\documentclass{article}
\RequirePackage{natbib}
\RequirePackage{amsthm,amsmath,latexsym,amssymb,graphicx}
\theoremstyle{plain}
\newtheorem{thm}{Theorem}

\begin{document}

\title{A note on conditional Akaike information for Poisson regression with random effects}

%\thankstext{T1}{Footnote to the title with the `thankstext' command.}

\author{Heng Lian\\Nanyang Technological University\\
Division of Mathematical Sciences\\
School of Physical and Mathematical Sciences\\
Nanyang Technological University\\
Singapore, 637371\\
}
\maketitle

\begin{abstract}
A popular model selection approach for generalized linear mixed-effects models is the Akaike information criterion, or AIC. Among others, \cite{vaida05} pointed out the distinction between the marginal and conditional inference depending on the focus of research. The conditional AIC was derived for the linear mixed-effects model which was later generalized by \cite{liang08}. We show that the similar strategy extends to Poisson regression with random effects, where condition AIC can be obtained based on our observations. Simulation studies demonstrate the usage of the criterion. 
\end{abstract}

\section{Introduction}
Generalized linear models (GLM) are powerful modelling tools that have gained popularity in statistics. It has wide applications in medical studies, pattern classification, sample surveys, etc. The scope of GLM can be greatly expanded by the incorporation of random effects. For example, in typical longitudinal studies, a model with random effects not only models individual characteristics, but attempts to extrapolate to the entire population as well. It takes into account both within cluster and between cluster variations in the study. Model selection in GLM is typically achieved using AIC or BIC combined with step-wise procedures. With fixed-effects models, the definition of AIC is straightforward using the likelihood penalized by a term that depends on the number of parameters. When random effects come into play, it is not entirely clear how the number of parameters in the model should be defined. Based on other previous works such as \cite{burnham98} and \cite{hodges01}, \cite{vaida05} made distinction between marginal and conditional inference and provided a formal definition of conditional Akaike information, cAI, which gives a theoretical justification for some previous approaches. They derived an unbiased estimator of cAI, called conditional AIC or cAIC, when the covariance matrix of random effects is known. \cite{liang08} derives a more general cAIC that dispenses with such strong assumptions. In \cite{vaida05}, the definition of cAI was given for general mixed-effects models but the unbiased estimator was only derived for linear mixed-effects models. A general approach of getting an unbiased estimator of cAI for generalized linear mixed-effects models (GLMM) seems to be out of reach. In this note, we propose an unbiased estimator of cAI for Poisson regression with random effects. The nature of Poisson regression is very different from the linear model since the responses are discrete. However, it turns out unbiased cAIC exists although it is derived in a different way. 

\section{Conditional AIC for count data}
Suppose we have some count responses $\{y_i\},i=1,\ldots, m$ from $m$ clusters that we want to model in relation to covariates $X_i$ and $Z_i$, with $y_i$ an $n_i\times 1$ vector from cluster $i$, and $X_i$,$Z_i$ are $n_i\times p$ and $n_i\times q$ matrices associated with fixed and random effects respectively. We use Poisson GLMM with the canonical link:
\begin{eqnarray}\label{model}
y_i&\sim& Pois(\mu_i)\\
\log\mu_i&=&X_i\beta+Z_ib_i,\; b_i\sim N(0, G),\nonumber
\end{eqnarray}
where $\beta$ is a $p\times 1$ vector of fixed effects and $b_i$ is a $q\times 1$ vector of random effects following a mean zero Gaussian distribution with unknown covariance matrix $G$. The total number of observations is thus $N=\sum_{i=1}^m n_i$. Let $\theta$ be the population parameters in the model, including $\beta$ and the parameters in $G$. The marginal likelihood is $g(y|\theta)=\int g(y|b,\theta)g(b|G)db$ where $g(y|b,\theta)$ is the Poisson likelihood conditional on the random effects and $g(b|G)$ is the density of the random effects. Sometimes it is more convenient to represent (\ref{model}) in the condensed form
\begin{eqnarray*}
y&\sim& Pois(\mu)\\
\log\mu&=&X\beta+Zb,
\end{eqnarray*}
where $y=(y_1^T,\ldots,y_m^T)^T$ is an $N\times 1$ vector of count responses, $X=(X_1^T,\ldots,X_m^T), Z=\mbox{diag}(Z_1,\ldots,Z_m)$ and $b=(b_1^T,\ldots,b_m^T)^T$.

In marginal inference, the focus is on the population parameters and the random effects are just a mechanism for modelling the correlations within the clusters. The standard AIC being used refers to this case and is called marginal AIC, mAIC, by \cite{vaida05}, defined by $-2\log g(y|\hat\theta(y))+2K$, where $K$ is the dimension of $\theta$. This penalty term is there to correct the bias caused by using the same data to estimate $\theta$ as well as to evaluate the marginal likelihood $g(y|\hat{\theta})$. The AIC is designed to approximate the Akaike information, $\mbox{AI}=-2E_{f(y)}E_{f(y*)}\log g(y^*|\hat{\theta})$, where $y^*$ is an independent replicate of $y$ coming from the same true distribution $f(y)$, which might not be contained within the family defined by (\ref{model}).

In conditional inference, the focus is on the cluster and the estimation of the random effects is of interest. The prediction in this case refers to new responses with the same clusters. Suppose the true distribution of $y$ is $f(y,u)=f(y|u)p(u)$ where $u$ is the true random effects with density $p(u)$. Following \cite{vaida05}, the conditional AI is naturally defined as 
\[\mbox{cAI}=-2E_{f(y,u)}E_{f(y^*|u)}\log g\{y^*|\hat{\theta}(y),\hat{b}(y)),\]
where $y^*$ is independent of $y$, generated from the same conditional distribution $f(\cdot|u)$. Similar to AI, cAI is cannot be directly calculated since the true distribution $f$ is unknown. For linear mixed-effects models, unbiased estimators were derived in \cite{vaida05} and \cite{liang08}. No unbiased estimator has been proposed for other GLMM to our best knowledge. The following theorem gives an unbiased estimator of cAI for Poisson regression, and the proof is given in the Appendix.
\begin{thm}
Assume that the count responses have the true distribution $y\sim Pois(\mu_0)$, where $\mu_0=(\mu_{01},\ldots, \mu_{0N}\}$ is the mean of the Poisson distribution and depends on some covariates as well as the random effects $u$. The data are modelled by (\ref{model}) with conditional likelihood denoted by $g(y|\theta,b)$. For any estimator $\hat{\theta}(y)$ and $\hat{b}(y)$, an unbiased estimator of the cAI is 
\[\mbox{cAIC}=-2\log g(y|\hat{\theta},\hat{b})+2K,\]
where $K$ is given by 
  \[\sum_{i=1}^N \left\{y_i\log[\hat{y}_i(y)]-y_i\log[\hat{y}_i(y^{y_i-1})]\right\}\]
and $y^{(y_i-1)}$ is the same as $y$ except its $i-$th component is replaced by $y_i-1$, and $y_i\log[\hat{y}_i(y^{(y_i-1)})]=0$ when $y_i=0$ by convention.
\end{thm}

Remark 1. Although the derivation of the unbiased estimator for cAI is different from the linear model, with the latter derived by integration by parts \citep{liang08}, the results have some resemblance with each other. For linear models, $K$ is given by $\sum_i\partial\hat{y}_i/\partial y_i$ and the partial derivatives are estimated by finite difference. Our $K$ for Poisson regression bares the similarity in that it depends on the difference between $\hat{y}_i$ and $\hat{y}_i(y^{(y_i-1)})$, the fitted responses after perturbing the original observations.

Remark 2. In Theorem 1, we only need to assume that the true model is in the Poisson family with means depending on some random effects $u$, which might also be different from the modelled random effects $b$. Thus the true model does not have to be included in the candidate model family. Besides, we are not assuming anything about the estimators $\hat{\theta}$ and $\hat{b}$ and they can be any reasonable estimators used in the literature. 
 
\section{Simulation study}
We conducted a simulation study to investigate the properties of our unbiased cAIC estimator and demonstrate the difference between marginal and conditional inference. We simulate data from model (\ref{model}) with a random intercept:
\begin{equation}
\log\mu_{ij}=\beta_0+\beta_1x_j+b_i,\; i=1,\ldots,m=10,\; j=0,\ldots,n_i,\nonumber
\end{equation}
where $\beta_0=1, \beta_1=0.2, x_j=j$ and $b_i\sim N(0,\sigma_b^2)$. In our simulation, we consider $n_i=5$ and $n_i=15$ with $\sigma_b=0.25, 0.5$ and $1$. For each of the six specifications, 500 data sets are generated. We compare the cAIC with the true bias, BC, defined by
\begin{equation}
\mbox{BC}=E_{f(y,u)}\log g(y|\hat{\theta},\hat{b})-E_{f(y,u)}E_{f(y^*|u)}\log g(y^*|\hat{\theta},\hat{b}),\nonumber
\end{equation}  
which is estimated by simulation with another independent $500$ sets of ${y^*}$'s generated from the true conditional distribution $f(\cdot|u)$ that shares common random effects $u$ with current responses.

The results are shown in Table 1. The estimated biases are close to the true value. In general, the `effective number of parameters' increases with the variance for the random effects. The same comparison can be made for mAIC and also for fixed-effects models, but we found in our simulations that the estimator $K$ in those cases is very close to the number of population parameters and there appears to be no advantages of using our estimator which only increases the computational burden. 
\begin{table}
\caption{Comparison of bias correction BC with its unbiased estimate, $K$, based on $500$ sets of simulated data}
\begin{tabular}{cccc}
\\
$n_i$&$\sigma_b$&BC&K\\
5  & 0.25&6.87&6.53\\
15 & 0.25&10.35&10.43\\
5  & 0.5 &9.18&9.09\\
15 & 0.5 &11.69&11.45\\
5  & 1   &10.19&10.31\\
15 & 1   &11.59&11.14\\
\end{tabular}
\end{table}

To illustrate the differences between  marginal and conditional inference, we use the same setup as before with $(\beta_0,\beta_1)=(1,0.2), n_i=5, \sigma_b=0.125, 0.25$ and $0.5$. Laplace approximation is used to approximate the marginal likelihood in the calculation of mAIC, for which the bias is simply estimated by the number of population parameters, $3$ in this case. Also, a fixed-effects model $\log\mu_{ij}=\beta_0+\beta_1x_j$ is fitted to the data and standard AIC is found. The values of AIC, mAIC and cAIC are shown in Figure 1 for different random effect variances. These values are averages over 500 sets of data simulated from the model. The difference between mAIC and cAIC is most obvious when $\sigma_b=0.125$. In fact, by comparing the information criteria, when $\sigma_b=0.125$, the fixed-effects model is selected for $395$ of the 500 data sets when comparing AIC with mAIC,  while it is selected only for $3$ data sets when comparing AIC with cAIC. When $\sigma_b=0.25$, fixed-effects model is selected for $165$ of the 500 data sets using mAIC, while it is selected only once when using cAIC.

\begin{figure}
\caption{Comparison of AIC (with fixed effects only), mAIC and cAIC}
\centerline{\includegraphics[width=6.5cm]{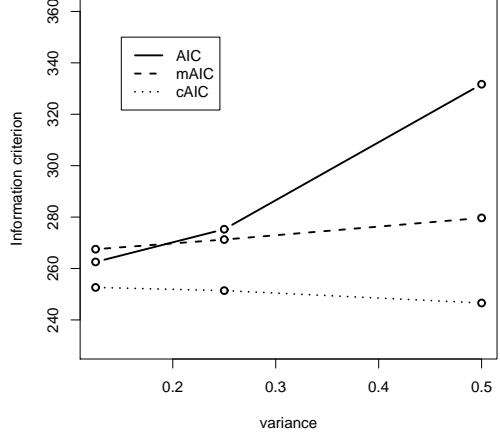}}
\end{figure}

\section{Concluding remarks}
Previous study of conditional AIC is only limited to the linear mixed-effects models. We provided the corresponding cAIC for Poisson regression. Since the derivation of the estimator does not depend on either the normality of random effects or specific estimators used for the fixed and random effects, the same formula works in more general contexts such as when using the approach of hierarchical likelihood \cite{lee96} which has become very popular in recent years. Although a general methodology seems to be lacking for generalized linear mixed-effects models, we believe that some approximation is possible and investigations in this direction are underway. 

\section*{Appendix}  
Proof of Theorem 1. Suppose that the true conditional likelihood is $f(y|u)=\prod_{i=1}^Ne^{-\mu_{0i}}\mu_{0i}^{y_i}/y_i!$, where $\mu_0=(\mu_{01}, \ldots, \mu_{0N})$ depends on the random effects $u$. Let $\hat{y}$ be the fitted responses from the mixed-effects model. The conditional Akaike information is 
\begin{eqnarray*}
\mbox{cAI}&=&-2E_{f(y,u)}E_{f(y^*|u)}\log g(y^*|\hat{\theta},\hat{b})\\
&=&-2E_{f(y,u)}E_{f(y^*|u)}\left[\sum_i(-\hat{y}_{i}+y^*_i\log\hat{y}_i-\log y^*_i!)\right]\\
&=&-2E_{f(y,u)}\left[\sum_i(-\hat{y}_{i}+\mu_{0i}\log\hat{y}_i-E_{f(y^*|u)}\log y^*_i!)\right].
\end{eqnarray*}
Meanwhile,
\begin{eqnarray*}
-2E_{f(y,u)}\log g(y|\hat{\theta},\hat{b})&=&-2E_{f(y,u)}\left[\sum_i(-\hat{y}_i+y_i\log\hat{y}_i-\log y_i!)\right].
\end{eqnarray*}
Thus
\[\mbox{cAI}-(-2E_{f(y,u)}\log g(y|\hat{\theta},\hat{b}))=2E_{f(y,u)}\sum_i (y_i-\mu_{0i})\log\hat{y}_i.\]
In addition, we have that
\begin{eqnarray*}
E_{f(y|u)}[\mu_{0i}\log\hat{y}_i]&=&\sum_{y_j,j\neq i}\sum_{y_i=0}^\infty\left\{\log[\hat{y}_i(y)]\mu_{0i}\frac{e^{-\mu_{0i}}\mu_{0i}^{y_i}}{y_i!}\prod_{j\neq i}\frac{e^{-\mu_{0j}}\mu_{0j}^{y_j}}{y_j!}\right\}\\
&=&\sum_{y_j,j\neq i}\sum_{y_i=0}^\infty\left\{(y_i+1)\log[\hat{y}_i(y)]\frac{e^{-\mu_{0i}}\mu_{0i}^{y_i+1}}{(y_i+1)!}\prod_{j\neq i}\frac{e^{-\mu_{0j}}\mu_{0j}^{y_j}}{y_j!}\right\}\\
&=&\sum_{y_j,j\neq i}\sum_{z_i=1}^\infty\left\{ z_i\log[\hat{y}_i(y^{(z_i-1)})]\frac{e^{-\mu_{0i}}\mu_{0i}^{z_i}}{z_i!}\prod_{j\neq i}\frac{e^{-\mu_{0j}}\mu_{0j}^{y_j}}{y_j!}\right\}\\
&=&\sum_{y_j,j\neq i}\sum_{z_i=0}^\infty\left\{z_i\log[\hat{y}_i(y^{(z_i-1)})]\frac{e^{-\mu_{0i}}\mu_{0i}^{z_i}}{z_i!}\prod_{j\neq i}\frac{e^{-\mu_{0j}}\mu_{0j}^{y_j}}{y_j!}\right\}\\
&=&E_{f(y|u)}\left\{y_i\log[\hat{y}_i(y^{(y_i-1)})]\right\},
\end{eqnarray*}
where $y^{(z_i-1)}$ is the vector $y$ whose $i-$th component has been replaced by $z_i-1$, and similarly for $y^{(y_i-1)}$.

Therefore, 
\begin{eqnarray*}
\mbox{cAI}-(-2E_{f(y,u)}\log g(y|\hat{\theta},\hat{b}))&=&2E_{p(u)}E_{f(y|u)}\left\{\sum_{i=1}^N (y_i-\mu_{0i})\log\hat{y}_i\right\}\\
&=&2E_{f(y,u)}\left\{\sum_{i=1}^N y_i\log\hat{y}_i-y_i\log[\hat{y}_i(y^{(y_i-1)})]\right\}.
\end{eqnarray*}
\bibliographystyle{plain}
\bibliography{caic.txt}

\end{document}